\documentclass[aip,amsmath,amssymb,preprint]{revtex4-1}
\usepackage{graphicx}
\usepackage{dcolumn}
\usepackage{bm}
\usepackage[mathlines]{lineno}

\usepackage[utf8]{inputenc}
\usepackage[T1]{fontenc}
\usepackage{mathptmx}
\usepackage{etoolbox}
\makeatletter
\def\@email#1#2{%
 \endgroup
 \patchcmd{\titleblock@produce}
  {\frontmatter@RRAPformat}
  {\frontmatter@RRAPformat{\produce@RRAP{*#1\href{mailto:#2}{#2}}}\frontmatter@RRAPformat}
  {}{}
}%
\makeatother
\begin{document}

\title{The effect of mechanical degradation on  vortex formation and decay in viscoelastic fluids}
\author{Renzo Guido, Luis G. Sarasua, Arturo C. Martí}
\affiliation{Facultad de Ciencias, Universidad de la Rep\'ublica, Igua 4225, Montevideo 11400, Uruguay}

\date{\today}

\begin{abstract}
Transient dynamics in viscoelastic fluids exhibits notable difference with their Newtonian counterpart. In this work we study the changes in the formation and decay of vortex of viscoelastic fluids due to degradation caused by shear stress. With this aim we performed long duration experiments with solutions of polyacrylamide confined between coaxial cylinders. 
The azimuthal velocity, acquired through digital particle velocimetry, shows oscillations before reaching a steady state. The vortex development is identified by the overshoot, which represents the difference between the maximum velocity and the stationary velocity. Similarly, during the vortex decay, the azimuthal velocity changes direction, and the relevant parameter is the undershoot, representing the maximum velocity of the transient reverse flow. The experimental results show that  overshoot and undershoot are closely related to the changes in the viscoelastic properties caused by the mechanical degradation. This effect opens new perspectives in the characterization of the viscoelastic fluids.
\end{abstract}
\maketitle



\section{Introduction}
\label{sec:sample1}

Flows induced by viscoelastic fluids hold significant academic and practical importance due to their involvement in numerous industrial and technological applications \cite{irgens}. Unlike Newtonian fluids, the rheological properties of polymer solutions are determined by a larger number of parameters \cite{Bird}. Furthermore, polymeric fluids are subject to the phenomenon of degradation, an irreversible transformation resulting from chemical or physical mechanisms \cite{soares2020review}. Mechanical degradation, specifically, occurs when the applied force exceeds the polymer chain's limit. Consequently, the chain breaks down leading to the formation of new polymers with reduced molecular weights. These alterations in the polymer structure engender changes in the rheological properties of viscoelastic fluids. In this study, we investigate the impact of mechanical stress on the transient dynamics of polymeric fluids. To accomplish this, we conducted long-duration experiments utilizing polyacrylamide solutions confined between coaxial cylinders, with a specific focus on analyzing the transient effects during the initiation and decay of vortical flows.

The transient dynamics of viscoelastics flows exhibits interesting contrasts with their Newtonian counterparts, behaving sometimes as expected and sometimes in a contra-intuitive way \cite{Shaqfeh,Qi,Fetecau2008,Khan2008,freire2019separation}. 
In our previous research \cite{guido2022development},  experimental 
results in cylindrical geometries revealed a  oscillatory behavior of the azimuthal velocity, acquired through the utilization of digital particle velocimetry, before reaching the stationary state. Inspired by  control  theory \cite{katsuhiko2010modern}, this phenomenon is characterized  by means of overshoot parameter, also known as maximum peak, which quantifies the relative difference between the maximum velocity and the stationary velocity.
Similarly, during the vortex decay, the azimuthal velocity undergoes a reversal in direction, and the pertinent parameter is referred to as undershoot, denoting the maximum negative velocity of the transient reverse flow.

The experimental results reported in \cite{guido2022development} revealed that the overshoot and undershoot variations are influenced by the elastic component and solvent viscosity. In the present work we show that the transient vortex dynamics in viscoelastic fluids can be related to mechanical degradation and present a model linking these phenomena. The rest of the paper is organized as follows.  In the next Section we present the experimental setup. Section~\ref{sec:results}
introduces the experimental results which are discussed in  Section~\ref{sec:discussion}. Finally, in Section~\ref{sec:conclusion} we present the conclusion.

\section{Experimental setup and working fluid}
\label{sec:setup}

\begin{figure}
\centering
\includegraphics[width=0.49\textwidth]{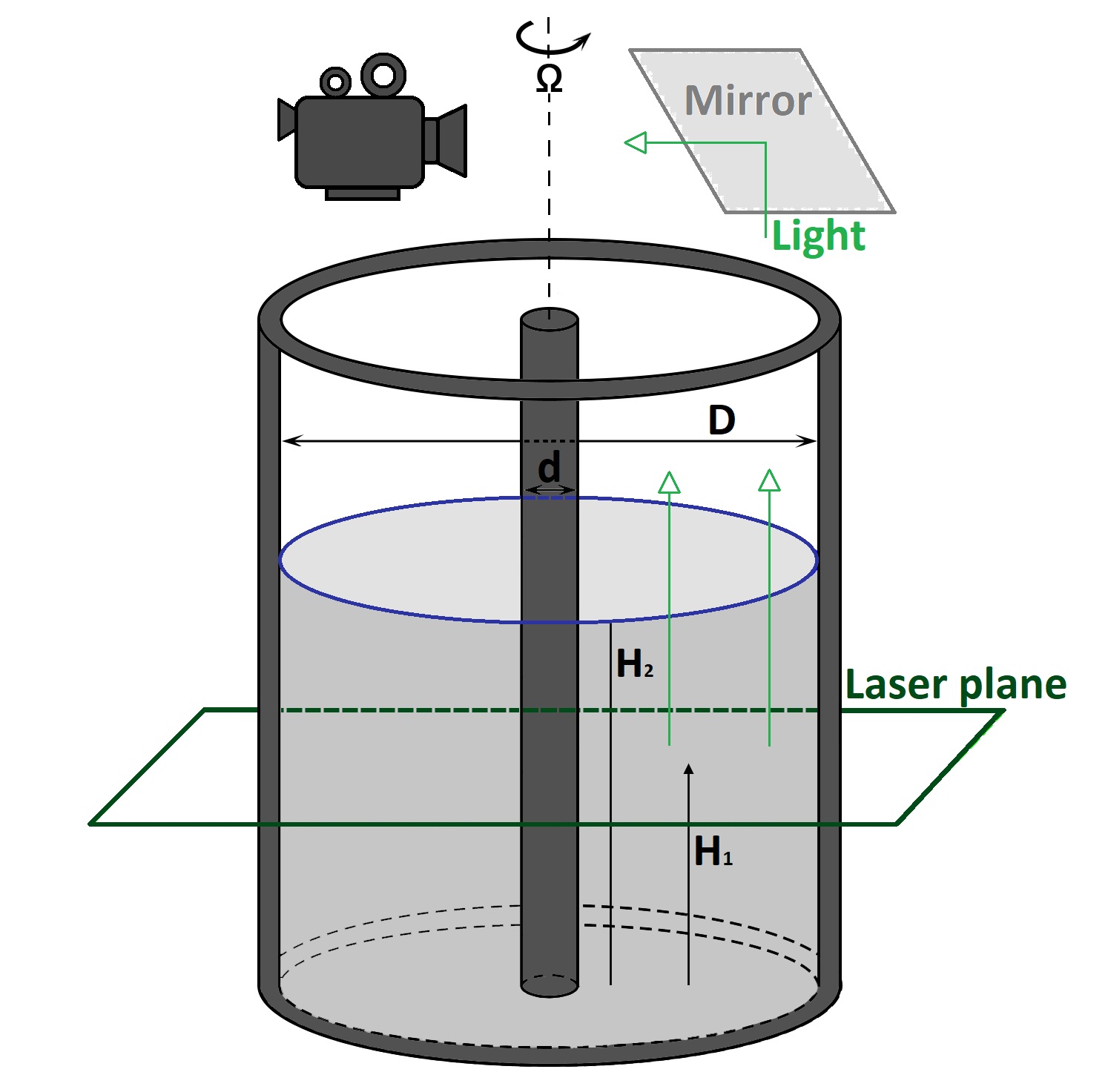}\includegraphics[width=0.49\textwidth]{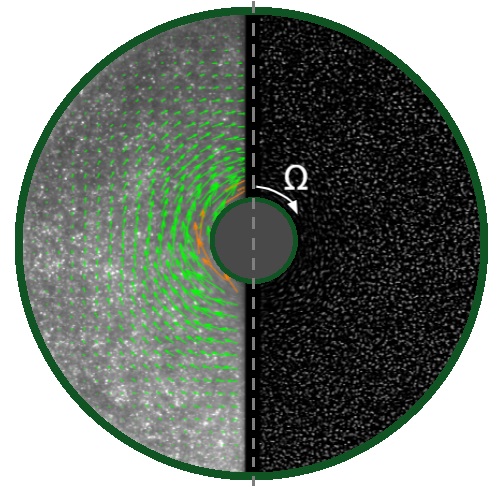}
\caption{The left panel shows a scheme of the experimental apparatus, the right panel is divided to depict an example of a typical snapshot of  the tracers and the velocity field obtained using the PIV technique.}
\label{fig:Aparato_Experimental}
\end{figure}

The experimental setup, schematized in Fig.~\ref{fig:Aparato_Experimental}, comprises two concentric cylinders. The outer cylinder ($D$ = 83.30 $\pm$ 0.05 mm) remains stationary, while the inner cylinder ($d$ = 9.50 $\pm$ 0.05 mm) is capable of rotation. The rotational velocity $\Omega$  of the inner cylinder is controlled by a direct current (dc) motor, capable of achieving velocities of up to 6.5 rad/s. The height of the fluid column is denoted as $h_1$ = 90 mm, and the laser sheet, positioned horizontally, is located at $h_2$ = 70 mm. The  bottom  lid is fixed, and the upper surface of the fluid is free, exposed to atmospheric pressure.

The fluid employed in this study was a shear-thinning solution of polyacrylamide (92560-50G Sigma-Aldrich)  in a mixture of water (70\%) and glycerin (30\%). The water, glycerin and polyacrylamide where mixed for 14 days with a magnetic stirrer Thermo Scientific (RT Basic-12) until the mixture was uniform. It is important to highlight that the properties of this mixture are not solely dependent on its composition but also influenced by the agitation time of the polyacrylamide.

\subsection{Degradation process and fluid characterization}
To investigate the effects of fluid degradation on its transient dynamics in our experiments, the fluid underwent additional agitation in the magnetic stirrer for 21 days. During this time, at days 5, 14 and 21, the fluid was placed in the cylindrical container to perform the start-up and decay of the rotating flow experiments. Samples of the fluid were taken at each stage to characterization, to measure the fluid properties changes due to degradation.    

The rheological properties of the samples were assessed using an Anton Paar Physica MCR 301 rheometer at a temperature of $20^{\circ}C$. Table \ref{tab:TablaReologia} presents the comprehensive summary of the fluid characteristics. 
The sample subjected to a degradation period of 5 days was not characterized, however, subsequent analysis demonstrates that its response closely resembles that of the original sample.
To characterize the rheological behaviour of each sample, the  Carrau mod\-el, expressed as $\eta (\dot{\gamma}) = \eta_0  [1+(\lambda \dot{\gamma})^2]^{(n-1)/2}$, was employed. By  fitting the data to this model, we obtained the  characteristic time $\lambda$ and the zero-shear rate  vis\-cos\-i\-ty $\eta_0$ for each sample. The Deborah number, 
defined as $De = \lambda U / R$, where $R$ represents the 
radius of the inner cylinder and $U$ denotes the velocity at 
its surface (thus $De = \lambda \Omega$), was computed. 
Additionally, the elasticity number, defined as $E = De/Re = 
\lambda \nu_0/R^2$, which is independent of the angular 
velocity, was calculated. Notably, both $\lambda$ and 
$\mu_0$ exhibited a decrease with mechanical degradation, 
attributed to the reduction in average chain length, 
consequently resulting in diminished elastic numbers.

\begin{table}
\centering
 \begin{tabular}{||c ||  c c c ||} 
 \hline
 Properties &  Original &  14 days & 21 days \\ [0.5ex] 
 \hline \hline
 $\mu_0$(Pa$\cdot $s) & $13.9$ & $7.8$ & $4.8$ \\ 
 \hline
 $\nu_0( \times 10^{-3}$ m$^2$/s) & $12.9$ & $7.23$ & $4.43$ \\  \hline
$\lambda$(s)  & $15$ & $7.3$ & $5.5$ \\ 
 \hline
 $\mathrm{Re}(\times 10^{-3})$  & $3.4-9.8$ &  $6.1-17.5$ & $9.8-28.4$ \\ 
 \hline
$\mathrm{De}$ & $29-84$ & $14-39$ & $11-31$ \\ 
 \hline
 $\mathrm{E} (\times 10^{3})$ & $8.6$ & $2.2$ & $1.1$ \\ [1ex]
 \hline
\end{tabular}
\caption{Fluid samples used along the experiments and their rheological properties. The glycerin content is $30\%$ and the polyacrylamide added was $2\%$ in weight. The rotational velocity used for calculating the Reynolds and Deborah number was the minimum (1.94 rad/s) and the maximum (5.60 rad/s).}
\label{tab:TablaReologia}
\end{table}

The acquisition of experimental velocity fields involved the introduction of neutrally buoyant, nearly-spherical polyamide particles (Dantec Dynamics, Denmark) into the viscoelastic fluid. These particles, with an average diameter of 50 $\mu$m, possessed a Stokes number of approximately $10^{-6}$, indicating their ability to closely trace the flow motion. The recording of the flow dynamics was accomplished through the utilization of a digital camera (Pixelink, model PL-B7760) operating at a frame rate of 15 \textit{fps}, facilitated by a mirror setup. The illumination source employed was a green laser (LaserGlow, model LSR-0532-PFH-005500-05n), with the particles reflecting its light as they followed the fluid motion.

\begin{figure}
\centering
\includegraphics[width=0.49\textwidth]{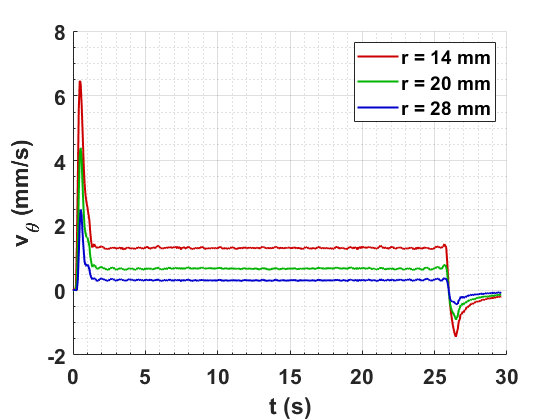}\includegraphics[width=0.49\textwidth]{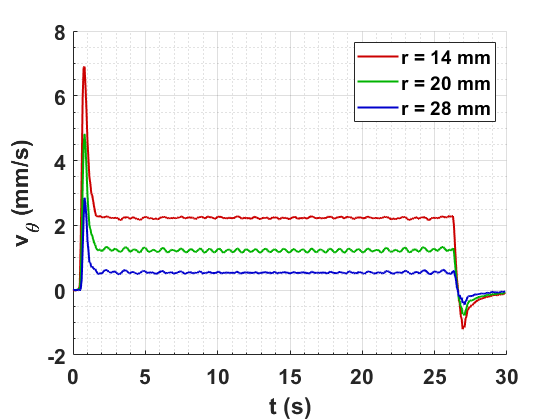}
\caption{Temporal evolution of the azimuthal velocity at different distances to rotation center indicated in the legend boxes. Left panel corresponds to the original fluid and the right panel to the same fluid after 14 days of degradation. In both cases the velocity of the inner cylinder is $4.8$ rad/s.}
        \label{fig:V_theta_t}
\end{figure}

For the measurement of the velocity field in Cartesian coordinates, the Digital Particle Imaging Velocimetry (DPIV) technique was employed. To determine the azimuthal component $v_{\theta}$, each grid point was categorized based on its distance from the rotation axis, partitioned into rings with a thickness of 2.05 mm. At each time instance, the average azimuthal velocity and its corresponding standard deviation were computed, yielding $v_{\theta} = v_{\theta}(r,t)$ as depicted in Fig.~\ref{fig:V_theta_t}.

The procedure in each experiment is as follows. The fluid is placed in the cylindrical container and allowed to stand for a reasonable period of time.  
The motor that controls the inner cylinder is started and quickly reaches a steady rotational speed denoted as $\Omega$. This sudden change in rotational velocity served as a step input. Subsequently, the fluid underwent the development of a vortex flow, characterized by a peak azimuthal velocity, followed by a gradual decay towards a stationary velocity field. After the fluid has developed the vortex flow we allow a short interval to elapse. Then, the motor is stopped and the inner cylinder ceased its rotation abruptly, leading to the decay of the vortex flow. 
We observe that the angular velocity of the fluid is decreasing but when it reaches the zero value instead of stopping completely changes the direction of rotation and for a few moments rotates in the opposite direction to the initial one. The time evolution of the velocity during this whole process is shown in the Fig.~\ref{fig:V_theta_t}.
This experimental procedure was systematically repeated, with each subsequent measurement featuring an increment in $\Omega$, ranging from $1.9$ to $5.6$ rad/s.

\section{Results}
\label{sec:results}

Consistent with our prior investigations, we conducted an analysis of the transient behavior in terms of the overshoot and undershoot  parameters. The overshoot, or maximum peak, was quantified using the expression $M_p = (v_{\theta \mathrm{max}} - v_{\theta \mathrm{st}} )/v_{\theta \mathrm{st}}$. It represents the relative difference between the maximum azimuthal velocity attained during the development of the vortex flow and the steady-state azimuthal velocity. Additionally, we assessed the undershoot, or, maximum negative peak, denoted as $M_{n p} = - v_{\theta \mathrm{min}} / v_{\theta \mathrm{st}}$. This parameter captures the maximum negative value of the azimuthal velocity  relative to the steady-state azimuthal velocity. Importantly, these parameters can be computed for each radial position and across various
angular velocities, $\Omega$, employed in our experiments.

\begin{figure}
\centering
\includegraphics[width=0.49\textwidth]{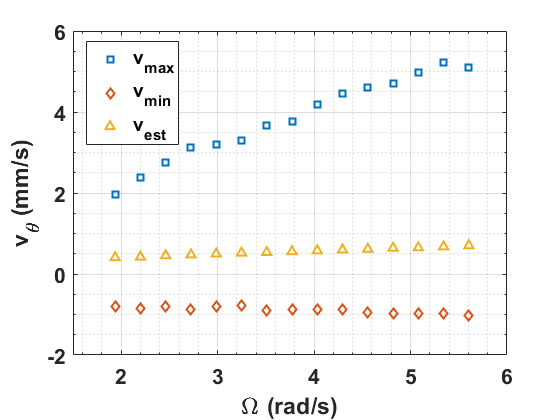}
\includegraphics[width=0.49\textwidth]{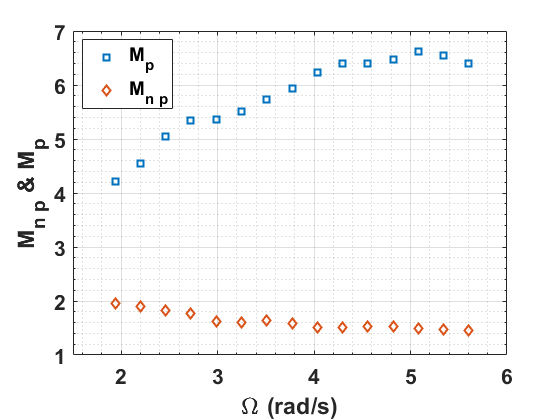}
\caption{Maximum, minimum and stationary azimuthal velocities (left panel) for the original  fluid  (without mechanical degradation) for $R$=$20$mm and  the corresponding overshoot and undershoot (right panel).}
\label{fig:SinDesgaste_Omega}
\end{figure}

In Fig.~\ref{fig:SinDesgaste_Omega}, the relationships between the maximum, minimum, and steady-state azimuthal velocities with the rotational velocity of the inner cylinder, $\Omega$, are depicted. The velocities exhibit an overall increase in magnitude as $\Omega$ increases. Additionally, the dimensionless parameters derived from the velocity data show that the overshoot parameter, $M_p$, becomes larger with increasing $\Omega$, indicating a greater deviation from the steady-state azimuthal velocity. In contrast, the undershoot parameter, $M_{n p}$, exhibits a negative dependence on $\Omega$, with its magnitude decreasing as the rotational velocity increases. These findings provide valuable insights into the complex flow behavior observed in the experimental system.

\begin{figure}
     \centering
         \centering
         \includegraphics[width=0.75\textwidth]{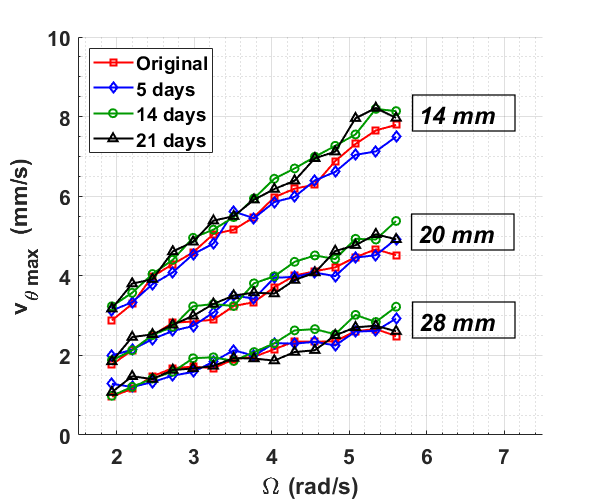}
         \label{fig:Vel_max_4fluidos}
        \caption{Maximum $v_{\theta}$ reached by each fluid at a fix radial coordinate as a function of $\Omega$. $v_{\theta max}$ appears to be the same disregarding the degradation for the same radial coordinate.}
        \label{fig:V_max_4fluidos}
\end{figure}

\begin{figure}
     \centering
         \centering
         \includegraphics[width=0.99\textwidth]{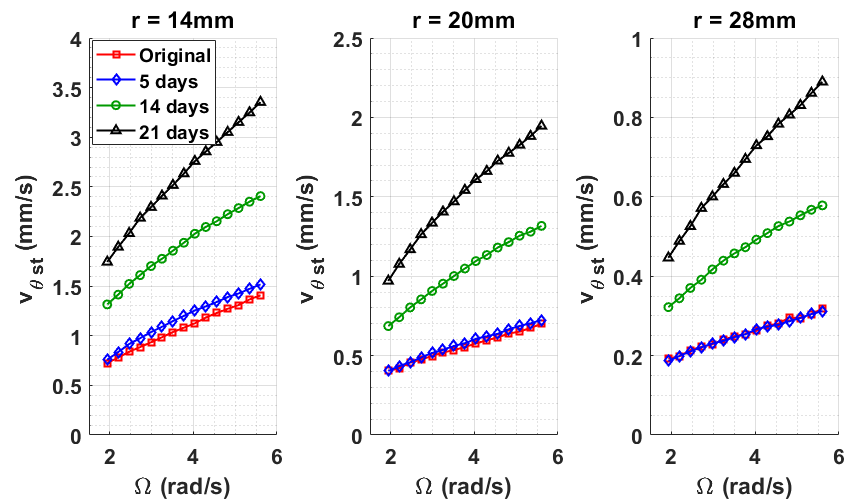}
         \label{fig:Vel_est_4fluidos}
        \caption{Stationary velocity $v_{\theta}$ reached by each fluid at the same radial coordinate as a function of $\Omega$. The degraded samples presented a larger stationary velocity, despite the 5 days sample that presented almost no change.}
        \label{fig:V_est_4fluidos}
\end{figure}

\begin{figure}
\centering
\includegraphics[width=0.99\textwidth]{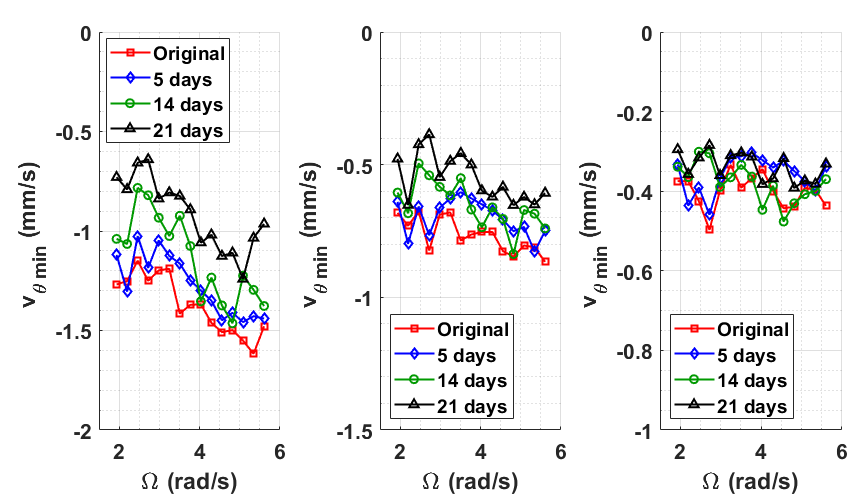}
\label{fig:Vel_min_4fluidos}
\caption{Minimum velocity $v_{\theta}$ reached by each fluid at
the same radial coordinate as a function of $\Omega$. The degradation samples presented a larger stationary velocity, so previous to the minimum the samples were not in the same state. The uncertainty of the measurement is larger than the stationary measurement.}
\label{fig:V_min_4fluidos}
\end{figure}

Next, we analyze  the maximum, stationary and minimum velocities  respectively for the fluid samples with four different degrees of degradation displayed in Figs.~\ref{fig:V_max_4fluidos}, \ref{fig:V_est_4fluidos} and \ref{fig:V_min_4fluidos}. The maximum velocity seems to be the same despite the degradation of the viscoelastic fluid, which can be interpreted as all samples exhibit initially the same response to the abrupt rotating start of the 
inner cylinder. This behaviour is notorious because the maximum peak is due to the elastic effects of the fluid, and although the elastic number decreases with degradation, the maximum velocity does not change considerably.

On the other side, in Figs.~\ref{fig:V_est_4fluidos} and \ref{fig:V_min_4fluidos} a notorious difference can be seen in both the stationary and minimum velocity for the different degrees of degradation. It can be noticed that the fluids response to 14 and 21 days of degradation is notorious, where increasing degradation increases stationary velocity. Nevertheless, the fluid response experiment almost no change after only five days of degradation. 
For the minimum velocity we found that the original fluid had a larger minimum velocity (in absolute value) compared to the others, despite the fact that it had a lower stationary velocity.

\section{Discussion}
\label{sec:discussion}

\begin{figure}
\centering
         \includegraphics[width=0.99\textwidth]{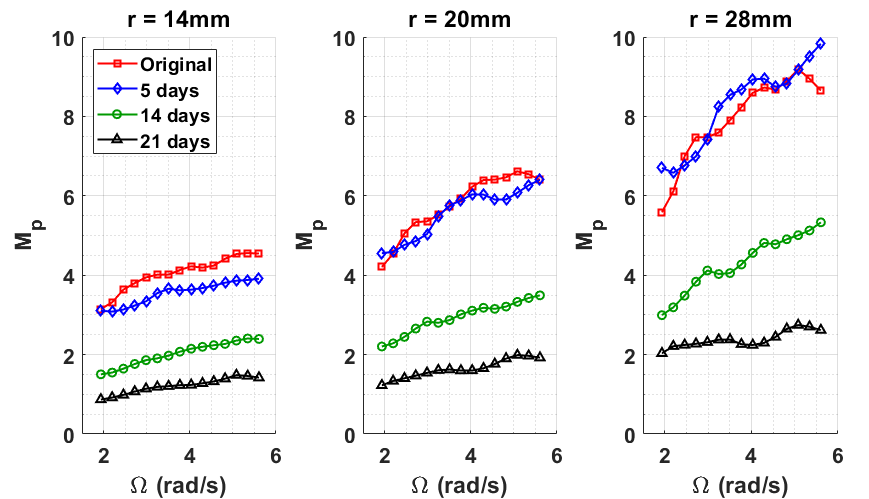}
         \caption{Maximum peak as a function of $\Omega$ for all fluids and for different  radial coordinates.}
        \label{fig:Os_4fluidos}
\end{figure}

\begin{figure}
     \centering
         \centering
         \includegraphics[width=0.99\textwidth]{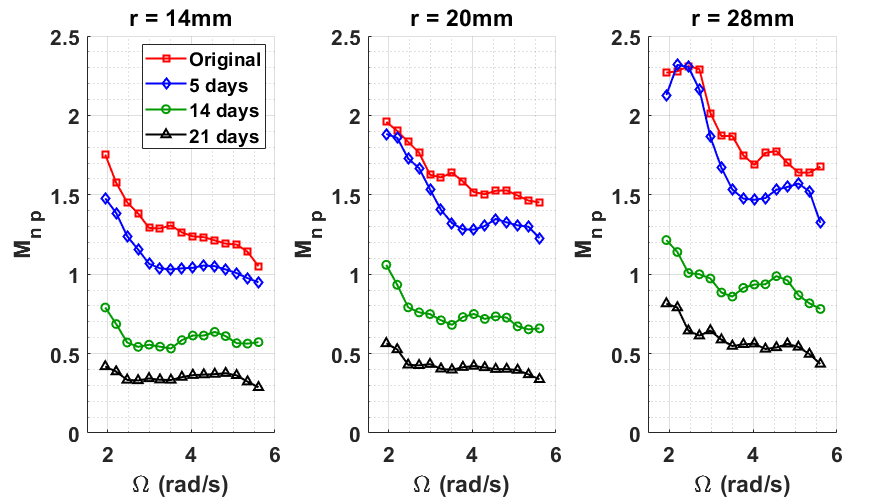}
                 \caption{Maximum negative peak as a function of $\Omega$ for all fluids and for different radial coordinates.}
        \label{fig:Us_4fluidos}
\end{figure}

To gain an understanding of vortex formation and decay, we calculated the previously defined dimensionless parameters as a function of the radial coordinate $r$  and  the rotational speed of the inner cylinder $\Omega$. The formation of the vortex is summarized in the overshoot parameter and we show it as a function of these parameters in Fig.~\ref{fig:Os_4fluidos}. As can be seen in the figure, the degradation shows a negative relation with the overshoot, and the original fluid and the 5-day degradation are almost identical. As expected, $M_p$ increases with $\Omega$ and  $r$.
On the other hand, the decay of the vortex is characterized by the $M_{n p}$,  plotted as a function of  the same parameters and different degrees of degradation in Fig.~\ref{fig:Us_4fluidos}. Again, the degradation presents a negative relation with the dimensionless parameter, and in this case $M_{n p}$ decreases with the distance $r$.

\begin{figure}
\centering
\includegraphics[width=0.49\textwidth]{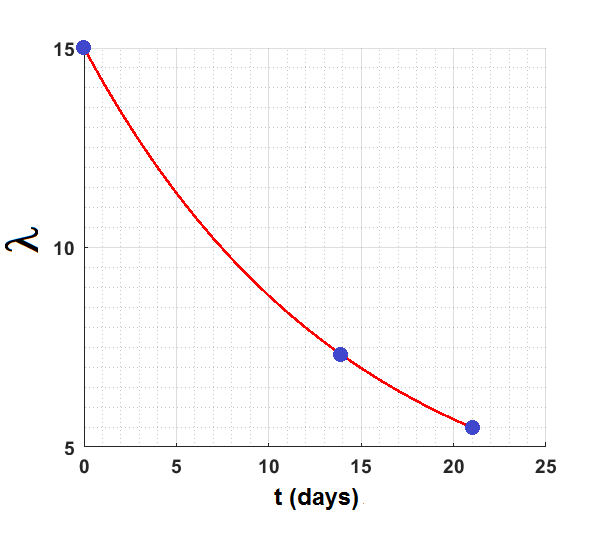}
\includegraphics[width=0.49\textwidth]{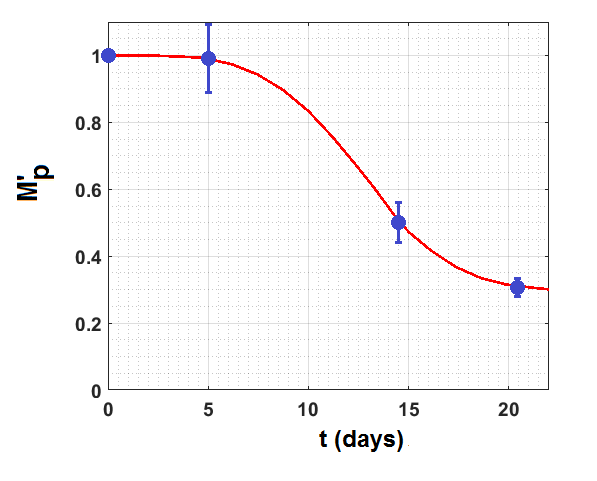}
\caption{The left panel  shows the decay of $\lambda$ in time due to degradation.  On the right hand side the decay of the average $M_p'$ in time is shown. The error bars indicate the dispersion.}
        \label{lambdat2}
\end{figure}

Next we study the parameters that characterize the transient dynamics of the vortex in relation to the changes in the rheological properties caused by the degradation process. As is shown in Fig.~\ref{lambdat2}, the decay in the characteristic time, $\lambda$, can be adequately described by an exponential function $\lambda(t)=A+B\exp(-(t/\tau_\lambda))$, with $\tau_\lambda=14.4$ days.
The  constant term $A$ in this expression indicates that, in agreement with the literature \cite{soares2020review}, the polymers are not totally degraded when the agitation is extended at constant  mixing regime. Instead, the length polymer distribution attains a stationary profile for which the most probable value is non-zero. This fact ensures that the elastic behavior does not disappear for  long times of agitation.

\begin{figure}
\centering
\includegraphics[width=0.49\textwidth]{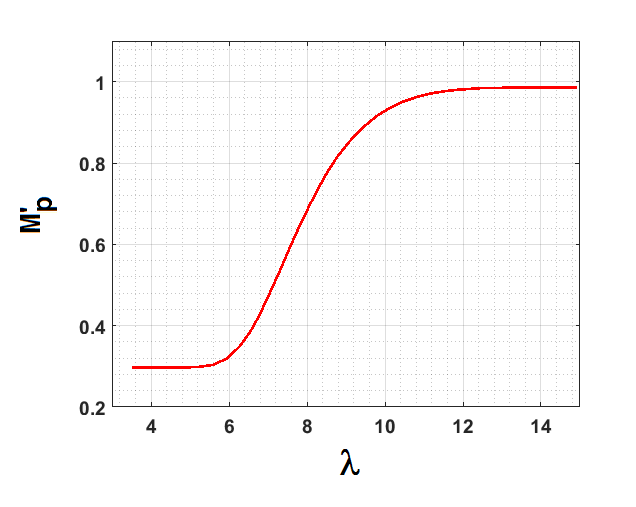}
 \includegraphics[width=0.49\textwidth]{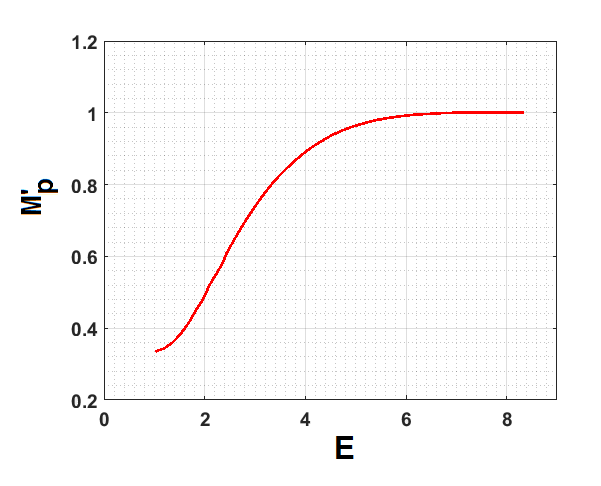}
\caption{Left panel: average $M_p'$ as a function of $\lambda$. Right panel: average $M_p'$ as a function of the elasticity number $E$.}
        \label{Mplambda2}
\end{figure}

On the other hand, we used the overshoot parameter to inspect the evolution of the flow.  It should be noted that the overshoot depends on the radial coordinate and the rotational velocity.  Then we considered the temporal 
average of the overshoot for different values of $r$ and $\Omega$. To do this we introduced a normalized overshoot $M_p'$, defined as $M_p'=M_p(t)/M_p(0)$, where  $M_p(0)$ represents the value of the overshoot at the initial stage.  The right panel of Fig.~\ref{lambdat2} shows the values of the averaged $M_p'$ as a function of time. 
As evidenced from this figure, the decaying of $M_p'$ for different values of $r$ and $\Omega$ are very similar. 
Since by definition $M_p = M_p(0) M_p'$ and $M_p(0)$ is a function of $r$ and $\Omega$, from the above considerations it follows that the overshoot can be written, for the considered range of parameters, as:
\begin{equation}
M_p(r,\Omega,t)=f(r,\Omega)M_p'(t) .   
\end{equation}
This expression is a consequence of the fact the evolution of $M_p'$ is almost the same for all the considered experiments. 
On the other hand, the evolution of the overshoot $M'_p$ cannot be described by a simple exponential function as is the case of the characteristic time. Thus we attempted to fit its variations to a function of the form $M_p'=C+D\exp(-(t/\tau_M)^a)$. The best  fit was obtained with $a=4$, $\tau_M=13.8$ days. Remarkably, the decay characteristic times $\tau_\lambda$ and $\tau_M$ obtained with the fits of $M_p'$ and $\lambda$ are very similar (the difference is less than 4$\%$). After  obtaining the decay in time of $M_p'$ and $\lambda$, we related these two quantities. The result is shown in the left panel of Fig.~\ref{Mplambda2}. As it can be seen, the value of $M_p'$ is weakly dependent on $\lambda$, for high values of $\lambda$, but it quickly reduces for small values of 
$\lambda$. This explains why the values of $M_p$ exhibit a slow decay at the initial times while decreasing faster for later times. 
We also considered the relationship between  $M_p'$ and the elasticity number $E$, which compares $De$ and $Re$. As mentioned previously, in the present system,   $E = \nu \lambda / R^2$. 

Similar to the previous procedure applied to the evolution of $\lambda$, we obtained the decay of $\nu$ from the rheology and then the values of $E$. In the right panel of Fig.~\ref{Mplambda2} we show the variation of $M_p'$ with $E$.  From this figure, it can be concluded that the overshoot is useful as a parameter to characterize the properties of the polymeric fluid.

\section{Conclusion}
\label{sec:conclusion}

We considered the transient dynamics of viscoelastic fluids, specifically, the formation and decay of vortex in cylindrical containers in terms of the mechanical degradation.
The experimental setup consists of two concentric cylinders, the outer one remains stationary while the inner one can be set in motion or stopped abruptly so that the fluid inside forms a vortex or decays respectively. The fluid dynamics differs from that of a Newtonian fluid since the azimuthal velocity exhibits oscillations. These oscillations are characterized by the overshoot during formation or the undershoot in the case of decay. Both parameters, overshoot and undershoot, allow characterizing this phenomenon.
    
We focused on the relationship between the rheological characteristic
of each sample, affected by the mechanical degradation resulting from 
the long stirring periods of several days
which lead to considerable alterations in the rheological properties
of the samples, and changes in the overshoot and undershoot.
From the experimental results, we observed a   decrease in elastic properties due to degradation. The drop on the elastic number in the fluid is due to changes in the medium length of the polyacrylamide chains, which presents a positive correlation with the Deborah number.  
It should be noted that, despite affecting the viscoelastic properties of the samples, the maximum azimuthal velocity does not change substantially with degradation time and the changes in overshoot are due to the changes in the stationary state velocity.

We obtained that the decay in the characteristic time, $\lambda$, can be adequately described by an exponential function $\lambda(t)=A+B\exp(-(t/\tau_\lambda))$, with $\tau_\lambda=14.4$ days.
The presence of the constant term $A$ in this expression indicates that, in agreement with the literature \cite{soares2020review}, the polymers are not totally degraded by agitation. As a consequence, the relaxation time does not tend to zero when $t \rightarrow \infty$. We also obtained that the overshoot parameter $M_p$ can be written as one function of time, that we called  normalized overshoot, and an amplitude that is a  function of $r$ and $\Omega$.
In light of these results, we determined a relationship between the overshoot and the elastic number. A potential application of this results would be to characterize viscoelastic fluid properties by means overshoot measurements.


 \bibliographystyle{elsarticle-num} 
 \bibliography{cas-refs}

\end{document}